\documentclass[prb]{revtex4}
\usepackage{amsmath}
\usepackage{graphicx}
\usepackage{amssymb}
\newcommand{\rr}[1] {\mathbf{r}_{#1 \parallel} }
\newcommand{\qq}[1] {\mathbf{q}_{#1 \parallel} }
\newcommand{\pp}[1] {\Pi^{(0)}_{#1}}
\begin{document}

\title{Comparison of charged particle energy loss in  epitaxial
with free-standing multilayer graphene.}
\author{O. Roslyak,$^1$   Godfrey Gumbs,$^1$ and Danhong Huang$^2$}
\affiliation{$^1$Department of Physics and Astronomy,
Hunter College at the City University of New York,
695 Park Avenue New York, NY 10065, USA}
\affiliation{$^2$Air Force Research Laboratory, Space Vehicles Directorate,\\
Kirtland Air Force Base, NM 87117, USA}

\date{\today}

\begin{abstract}
We present a formalism and numerical  results  for the energy loss of a charged
particle scattered at an arbitrary angle from  epitaxially grown multilayer graphene
(MLG). It is compared with that of free-standing graphene layers. Specifically, we
investigated the effect of the substrate induced energy gap on one of the layers.
The gap yields collective plasma oscillations whose characteristics are qualitatively
and quantitatively different from those produced by Dirac fermions in gapless graphene.
The range of wave numbers for undamped self-sustaining plasmons is increased as the
gap is increased, thereby increasing and red-shifting the MLG stopping power for some range of
charged particle velocity. We also applied our formalism to interpret several distinct features of
experimentally obtained electron energy loss spectroscopy (EELS) data.
\end{abstract}

\pacs{73.21.-b, 03.67.Lx, 71.70.Ej}

\maketitle

\section{Introduction}
\label{sec1}

\par
Electron energy loss spectroscopy (EELS) has been considered for many systems dating
back to the classic paper of Ritchie\,\cite{ritchie}  on surface plasmons
for a slab of dielectric material in the local limit and subsequently generalized
to a non-local dielectric function by Gumbs and Horing.\,\cite{slab} The non-locality
was included in Ref.\,\onlinecite{slab} through the screening of the electron-electron
interaction produced by single-particle excitations and plasmon modes. Fessatidis,
et al.\,\cite{fessatidis} recently adopted the formalism of Horing, et al.\,\cite{tso}
for a two-dimensional electron gas (2DEG) and Gumbs and Balassis\,\cite{GG} for a nanotube
to  graphene with the aid of the polarization function calculated by
Wunsch, et al.\,\cite{wunsch} for conventional Dirac electrons.

\par
There have been several recent papers which considered  the effects which a
circularly polarized electromagnetic field (CPEF),\,\cite{kibis,roslyak}
spin-orbit interaction (SOI)\,\cite{SOI} in suspended graphene or the sublattice symmetry breaking (SSB)
by an underlying polar substrate\,\cite{SOI-2,li_2009,giovannetti_2007} in epitaxial
graphene may have on the energy band structure and plasma excitations\,\cite{roslyak}  of
a graphene sheet. Under these conditions, there is a gap  between the valence and
conduction bands as well as between the intra-band and inter-band electron-hole
continuum of the otherwise semimetallic Dirac system.\,\cite{wunsch,shung1,shung2}
Additionally, the interplay between the single-particle excitations in the long wavelength
limit results in dielectric screening of the Coulomb interaction which produces
an undamped plasmon mode that appears in the gap separating the two
types of electron-hole modes forming a continuum. By this we mean that a
self-sustained collective plasma mode is not supported by exciting \emph{either}
the intra-band or inter-band single-particle modes only. Furthermore,
since the Dirac electrons now acquire a non-zero effective
mass, one can produce a long wavelength plasmon mode as in the
2DEG by intra-band excitations only. These
properties of the plasma modes give rise to noticeable differences in the behavior
of the stopping power of gapped graphene compared with conventional
free-standing exfoliated multilayer graphene.

\par
Multi-layer epitaxially grown graphene (MLG) may become a valuable and relatively cheap
alternative to rather expensive exfoliated graphene.
Recent angle-resolved photoemission spectroscopy (ARPES) experiments unambiguously demonstrated almost perfect Dirac cones on most of the layers.
That is the electrons in the layers behave as if the layers are uncoupled in
contrast to Bernal stacking in bi-layer graphene.\,\cite{SPRINKLE:2010}
Although some layers may establish bi-layer structures the interaction
between the first and the buffer layer (the one sitting on top of the SiC substrate)
is always weak.\,\cite{MATTAUSCH:2007,MALLET:2007}
The spectral function of MLG on SiC substrate suggests
the energy gap of few hundreds of $\texttt{meV}$.
However the exact gap opening mechanism in epitaxial graphene is still under debate.\,\cite{rotenberg_2008}
The complexity comes from the fact that the graphene
sample  sits on top of a buffer layer which provides an additional mid gap
level, thus obscuring the exact energy dispersion curve and
requires numerical \emph{ab initio} calculations.\,\cite{kim_2008}
In addition, the DOS around Fermi energy failed to indicate the gap.\,\cite{SPRINKLE:2010}
This ambiguity stimulated  discussion regarding the symmetry breaking
gap\,\cite{zhou_2007} versus the effects due to  electron-electron
interaction.\,\cite{bostwik_2007} This also indicates the importance
of alternative techniques capable of identifying the induced gap by a buffer layer on a substrate.

\par
EELS  may be employed to  ascertain
the plasmon frequencies in single and double layer graphene.
The Raman shift of the scattered electrons provides both particle-hole
and plasmon excitation frequencies, which are usually characterized by
their spectral weight, a quantity that  depends on the
transferred energy $\hbar\omega$ and in-plane momentum $\hbar q$.
This allows mapping of the plasmon dispersions $\omega_p(q)$.\,\cite{LU:2009}
Here we calculate EELS spectra in order to pinpoint the effect of the
substrate induced gap in MLG.

\par
The outline of the  remainder of this paper is as follows.
In Sec.\,\ref{SEC:2}, we derive the energy loss for charged
particles incident on multiple electron layers at an arbitrary angle of incidence.
In Sec.\,\ref{SEC:3}, this  formalism is applied to MLG with a gapped
buffer layer. Our numerical simulations are compared with experimental results
of Lu and Loh\,\cite{LU:2009} in Sec.\,\ref{SEC:4}.
A summary is presented in Sec.\,\ref{SEC:5}.

\section{Energy Loss Formalism}
\label{SEC:2}

We shall consider an inhomogeneous medium (in the $\hat{\mathbf{z}}$ direction)
described by a nonlocal dielectric function $\epsilon \left({1;2}\right)$. Here,
we have introduced space-time points, such as $i = \left({z_i, \mathbf{r}_i, t_i}\right)$,
where the $z$ spacial coordinate has been separated out due to the inhomogeneity and
making cylindrical coordinates suitable for our discussion. Our focus is the calculation
of energy  loss of a charged particle $Z e$ moving along a  prescribed trajectory
$z_1 + \mathbf{v} t_1$. Due to non-locality of the dielectric function, the charged particle
experiences a  frictional force given by the gradient of the effective potential as

\begin{equation}
\label{EQ:3_1}
\mathbf{F}(1) = - Z e \left[{\nabla_1 \Phi_{\texttt{tot}}(1)}\right]_{1=z_1 + \mathbf{v} t_1} \ .
\end{equation}
The total potential can be determined from the following system of equations

\begin{gather}
\label{EQ:3_2}
\Phi_{\texttt{tot}}(1) = \int d^4 2\  \epsilon^{-1}(1;2) \Phi_{\texttt{ext}}(2)\ ,\\
\label{EQ:3_3}
\nabla^2_1 \Phi_{\texttt{ext}} (1) = -\frac{Ze}{\epsilon_0}\,\delta (1-z_1 - \mathbf{v} t_1) \ .
\end{gather}
Solving Eq.\ \eqref{EQ:3_3} by Fourier transformation and  substituting
the result into Eq.\,\eqref{EQ:3_2} and then into \eqref{EQ:3_1},
we obtain the components of the frictional force as

\begin{gather}
\label{EQ:3_4}
\mathbf{F}_{\parallel} (1) = - \frac{\left({Z e}\right)^2}{\left({2 \pi}\right)^3\epsilon_0}
\int \limits_{-\infty}^{\infty} d z_2
\int  i \mathbf{q_{\parallel}}\,d^2\  \mathbf{q_{\parallel}}
\int \limits_{-\infty}^{\infty} d q_z\,
  \texttt{e}^{i q_z \left({z_2 - z_1 -  v_z t_1}\right)}\,q^{-2}
\epsilon^{-1} \left({z_1+ v_z t_1,z_2; q_{\parallel}, \mathbf{q} \cdot \mathbf{v}}\right)\ ,
\\
\label{EQ:3_5}
\mathbf{F}_{\perp}(1)  =  - \frac{\left({Z e}\right)^2}{\left({2 \pi}\right)^3\epsilon_0}
\int \limits_{-\infty}^{\infty} d z_2
\int d^2 \mathbf{q_{\parallel}}
\int \limits_{-\infty}^{\infty} d q_z  \texttt{e}^{i q_z \left({z_2 - z_1 - v_z t_1}\right)}\,q^{-2}
\frac{\partial}{\partial \left({z_1+ v_z t_1}\right)}\,\epsilon^{-1}
\left({z_1+ v_z t_1,z_2; q_{\parallel}, \mathbf{q} \cdot \mathbf{v}}\right)  \ .
\end{gather}
We assume that the particle trajectory begins  at  $t=-\infty$ and ends at
$t=\infty$ with its motion in  vacuum. The net energy lost due to the frictional
force between  it and the plasma  during this motion may be expressed as

\begin{equation}
\label{EQ:3_6}
W = \int \limits_{-\infty}^{\infty} dt_1\,\mathbf{v} (t_1) \cdot \mathbf{F}(1) \ .
\end{equation}
It is convenient to partition the loss as $W = W_{\parallel} + W_{\perp}$. The energy
loss due to parallel and perpendicular motions may be written as

\begin{gather}
\label{EQ:3_71}
W_{\parallel} \left({z_1, \mathbf{v}}\right) = -\frac{i\left({Z e}\right)^2}{\left({2 \pi}\right)^3\epsilon_0}
\int \limits_{-\infty}^{\infty} d t_1
\int \limits_{-\infty}^{\infty} d z_2
\int  d^2 \mathbf{q_{\parallel}} \    \left({\mathbf{q_{\parallel}} \cdot \mathbf{v}_{\parallel}}\right)
\int \limits_{-\infty}^{\infty} d q_z\
  \texttt{e}^{i q_z \left({z_2 - z_1 - v_z t_1}\right)}\,q^{-2}
\epsilon^{-1} \left({z_1+ v_z t_1,z_2; q_{\parallel}, \mathbf{q} \cdot \mathbf{v}}\right)\ ,\\
\label{EQ:3_72}
W_{\perp} \left({\mathbf{v}}\right) =  -\frac{i\left({Z e}\right)^2}{\left({2 \pi}\right)^3\epsilon_0}
\int \limits_{-\infty}^{\infty} d z_1
\int \limits_{-\infty}^{\infty} d z_2
\int  d^2 \mathbf{q_{\parallel}}
\int \limits_{-\infty}^{\infty}   dq_z \ q_z  \texttt{e}^{i q_z \left({z_2 - z_1}\right)}\,q^{-2}
\epsilon^{-1} \left({z_1,z_2; q_{\parallel}, \mathbf{q} \cdot \mathbf{v}}\right)\ .
\end{gather}
In Eq.\,\eqref{EQ:3_72}, integration by parts was carried out, along with the condition
that the charged particle starts and ends in   vacuum:
$\epsilon^{-1} \left({-\infty,z_2; q_{\parallel}, \mathbf{q} \cdot \mathbf{v}}\right) = \epsilon^{-1} \left({\infty,z_2; q_{\parallel}, \mathbf{q} \cdot \mathbf{v}}\right) = 1$.
The above general expressions can be simplified further, for the practically important cases of parallel,
perpendicular and traversing motion.

\subsection{Charged Particle moving parallel to the surface   ($\mathbf{v}_{z} = 0$)}

First, we assume that the charged particle   travels parallel to the $x,y$ plane
with velocity component  $v_z = 0$.
In this case, to simplify Eq.\,\eqref{EQ:3_71}, we can use the following identities:

\begin{gather}
\label{EQ:3_8}
\int \limits_{-\infty}^{\infty}  d q_z\ q^{-2}  \texttt{e}^{i q_z \left({z_2 - z_1}\right)} = 2 \pi \frac{\texttt{e}^{-q_{\parallel} \vert{z_1 - z_2}\vert}}{2 q_{\parallel}}\ ,\\
\label{EQ:3_9}
i \int   \left({\mathbf{q_{\parallel}} \cdot \mathbf{v}_{\parallel}}\right) d^2 \mathbf{q_{\parallel}}
\epsilon^{-1} \left({z_1,z_2; q_{\parallel}, \mathbf{q_{\parallel}} \cdot \mathbf{v_{\parallel}}}\right)
= - 2 i \int \limits_{0}^{\infty} v_{\parallel} d q_{\parallel}\ q^2_{\parallel}
\int \limits_{-1}^{1} \frac{\cos \theta\  d(\cos \theta)}{\sqrt{1 - \cos^2 \theta}}
\epsilon^{-1} \left({z_1,z_2; q_{\parallel}, q_{\parallel} v_{\parallel} \cos \theta}\right)\\
\notag
=
- \int  d^2 \mathbf{q_{\parallel}}\  \left({\mathbf{q_{\parallel}} \cdot \mathbf{v}_{\parallel}}\right)
\Im \texttt{m} \epsilon^{-1} \left({z_1,z_2; q_{\parallel},
\mathbf{q_{\parallel}} \cdot \mathbf{v_{\parallel}}}\right) \ .
\end{gather}
In the above Eq.~\eqref{EQ:3_9}, to integrate over the angle we   used the
parity of the inverse  dielectric function, i.e.,

\begin{equation}
\label{EQ:3_10}
\epsilon^{-1} \left({z_1,z_2;-\mathbf{q}_{\parallel},-\omega}\right)
= \left[\epsilon^{-1} \left({z_1,z_2;\mathbf{q}_{\parallel},\omega}\right)\right]^\ast\ .
\end{equation}
Consequently, we obtain the energy loss in the parallel case as

\begin{equation}
\label{EQ:3_11}
W_{\parallel} \left({z_1, v_{\parallel}}\right) = \left({\frac{Z e}{2 \pi}}\right)^2
\int \limits_{-\infty}^{\infty} d t_1
\int \limits_{-\infty}^{\infty} d z_2
\int  d^2 \mathbf{q_{\parallel}}\
 \left({\mathbf{q_{\parallel}} \cdot \mathbf{v}_{\parallel}}\right)
\frac{\texttt{e}^{-q_{\parallel} \vert{z_1 - z_2}\vert}}{2\epsilon_0q_{\parallel}}
\Im \texttt{m} \epsilon^{-1} \left({z_1,z_2; q_{\parallel}, \mathbf{q_{\parallel}}
\cdot \mathbf{v_{\parallel}}}\right) \ .
\end{equation}
Since the integrand in the above expression does not depend on time, it is more convenient
to represent it in the form of constant energy loss rate (per unit time) as

\begin{gather}
\label{EQ:3_12}
W_{\parallel} \left({z_1, v_{\parallel}}\right) =  \int \limits_{-\infty}^{\infty}
d t_1 \frac{d W_{\parallel}}{dt}\\
\notag
\frac{d W_{\parallel}}{dt} \left({z_1, v_{\parallel}}\right)  =
- \left({\frac{Z e}{2 \pi}}\right)^2
\int \limits_{-\infty}^{\infty} d z_2
\int d^2 \mathbf{q_{\parallel}}\
  \left({\mathbf{q_{\parallel}} \cdot \mathbf{v}_{\parallel}}\right)
\frac{\texttt{e}^{-q_{\parallel} \vert{z_1 - z_2}\vert}}{2\epsilon_0q_{\parallel}}
\Im \texttt{m} \epsilon^{-1} \left({z_1,z_2; q_{\parallel}, - \mathbf{q_{\parallel}} \cdot \mathbf{v_{\parallel}}}\right) \ .
\end{gather}
In the above, we used the odd parity of the imaginary part of the inverse dielectric
function. The rate of the energy loss expressed per unit length
is usually referred to  as the stopping power $S = \frac{d W_{\parallel}}{v_{\parallel}dt}$.

\subsection{Charged Particle moving perpendicular to the
surface ($\mathbf{v}_{\parallel} = 0$)}

Now, let us assume that the particle has only velocity component perpendicular
to the $x,y$ plane, i.e., $v_{\parallel} = 0$.
We shall focus on the last integration in Eq.\,\eqref{EQ:3_72}, by changing the
variable $\omega = q_z v_z$, so that we can perform the
series string of simplifications on it, namely,

\begin{gather}
\label{EQ:3_13}
\frac{1}{v_z^2} \int \limits_{-\infty}^{\infty}  d \omega\ \omega
 \frac{\texttt{e}^{i \omega \left({z_2- z_1}\right)/v_z}}{q_{\parallel}^2 +
\left({\omega/v_z}\right)^2} \epsilon^{-1}\left({z_1,z_2; q_{\parallel},\omega}\right)
\\
\notag
= \frac{1}{v_z^2} \int \limits_{0}^{\infty}  d \omega\ \omega
 \frac{\texttt{e}^{i \omega \left({z_2- z_1}\right)/v_z}}{q_{\parallel}^2 +
\left({\omega/v_z}\right)^2} \epsilon^{-1}\left({z_1,z_2; q_{\parallel},\omega}\right) -
\frac{1}{v_z^2} \int \limits_{0}^{\infty}  d \omega\ \omega
 \frac{\texttt{e}^{-i \omega \left({z_2- z_1}\right)/v_z}}{q_{\parallel}^2 +
\left({\omega/v_z}\right)^2} (\epsilon^\star)^{-1}\left({z_1,z_2; q_{\parallel},\omega}\right)   \\
\notag
=  \frac{2 i }{v_z^2} \int \limits_{0}^{\infty}
\frac{  d \omega\ \omega }{q_{\parallel}^2 +
\left({\omega/v_z}\right)^2} \Im \texttt{m} \; \texttt{e}^{i \omega \left({z_2- z_1}\right)/v_z} \epsilon^{-1}\left({z_1,z_2; q_{\parallel},\omega}\right)\ .
\end{gather}
Consequently, by combining Eqs.\,\eqref{EQ:3_13} and \eqref{EQ:3_72}, we
obtain the following energy loss in the perpendicular case as

\begin{gather}
\label{EQ:3_14}
W_{\perp} \left({v_z}\right) =   \frac{2\left({Z e}\right)^2}{\left({2 \pi}\right)^3\epsilon_0}
\Im \texttt{m} \;
\int \limits_{-\infty}^{\infty} d z_1
\int \limits_{-\infty}^{\infty} d z_2
\int  d^2 \qq{}
\int \limits_{0}^{\infty}\frac{  d\omega\ \omega}{
\left({q_{\parallel} v_z}\right)^2+\omega^2}
\epsilon^{-1}\left({z_1,z_2; q_{\parallel},\omega}\right)
\texttt{e}^{i \omega \left({z_2- z_1}\right)/v_z}\ .
\end{gather}
Now, let us apply the energy loss formalism to single and double layers of
of epitaxially grown graphene. The epitaxial  form of graphene usually
features the energy gap, which may be controlled by an external electric field
from a gate.

\subsection{Charged Particle traversing/reflected from the graphene layers at an arbitrary angle }

For the  case when the charged particle crosses the layers, none of the
components of the velocity is zero. As in the perpendicular case, the initial
position of the particle is of little importance and we may change
variable $z_1 + v_z t \to z_1$ in Eq.\,\eqref{EQ:3_71}, followed by adding
the obtained equation to Eq.\,\eqref{EQ:3_72}, giving

\begin{gather}
\label{EQ:3_15}
W_{\theta}\left({\mathbf{v}}\right)=
 - \frac{i\left({Z e}\right)^2}{\left({2 \pi}\right)^3 v_z\epsilon_0}
\int \limits_{-\infty}^{\infty} d z_1\,\int \limits_{-\infty}^{\infty} d z_2\,\int \limits_{-\infty}^{\infty}d q_z \int d^2\,
 \mathbf{q}_{\parallel}
\texttt{e}^{i q_z \left({z_2 - z_1}\right)}
\epsilon^{-1}\left({z_1,z_2; q_{\parallel}, \mathbf{q} \cdot \mathbf{v}}\right) q^{-2}
\mathbf{q} \cdot \mathbf{v}  \\
\notag
= \frac{2\left({Z e}\right)^2}{\left({2 \pi}\right)^3 v_z\epsilon_0} \Im \texttt{m} \;
\int \limits_{-\infty}^{\infty} d z_1\,\int \limits_{-\infty}^{\infty}
d z_2\,
\int \limits_{0}^{\infty} dq_z\,\int d^2 \mathbf{q}_{\parallel}\,
\texttt{e}^{i q_z \left({z_2 - z_1}\right)}
\epsilon^{-1}\left({z_1,z_2; q_{\parallel}, \mathbf{q} \cdot \mathbf{v}}\right) q^{-2} \mathbf{q} \cdot \mathbf{v} \ .
\end{gather}
The last expression is obtained by changing the sign $\mathbf{q} \to -\mathbf{q}$
and utilizing the symmetry relation Eq.\,\eqref{EQ:3_10}.
We now  rewrite the energy loss via the energy-loss probability spectral
function $\mathcal{P}\left({\omega}\right)$.
To do so, we introduce $\omega = \mathbf{q} \cdot \mathbf{v} = q_z v_z + \qq{} \cdot \mathbf{v}_{\parallel}$, thereby obtaining

\begin{gather}
\label{EQ:3_15_1}
W_{\theta}\left({\mathbf{v}}\right)= \int \limits_{0}^{\infty} d \omega\ \omega
\mathcal{P} \left({\omega,\mathbf{v}}\right)\\
\notag
\mathcal{P}\left({\omega,\mathbf{v}}\right) =
\frac{2(Z e)^2 }{\left({2 \pi}\right)^3\epsilon_0}
\Im \texttt{m}
\int \limits_{-\infty}^{\infty} d z_1 \int \limits_{-\infty}^{\infty} d z_2
\int d^2 \qq{}\
\epsilon^{-1} \left({z_1,z_2; \qq{},\omega}\right)
\frac{\exp\left({ i \left({\omega - \mathbf{q}_{\parallel}
\cdot \mathbf{v}_{\parallel}}\right)(z_2-z_1)/v_z}\right)}
{\left({q_{\parallel} v_z}\right)^2 +
\left({\omega - \mathbf{q}_\parallel}
\cdot \mathbf{v}_{\parallel}\right)^2} \ .
\end{gather}
An alternative derivation of the above equation is provided in Appendix\,\ref{AP:1}.
One may show in a straightforward way that   the above general expression may
be expressed as Eq.\,\eqref{EQ:3_14} in the case of $\mathbf{v}_{\parallel} = 0$.
The other limiting case Eq.\,\eqref{EQ:3_12} is less trivial.
It is obtained from Eq.\,\eqref{EQ:3_15_1} by taking the following limits
$\lim \limits_{v_z \rightarrow 0} \left({\omega - \qq{} \cdot \mathbf{v}_{\parallel}}\right)/v_z = q_z$,
$\lim \limits_{v_z \rightarrow 0} d z_1/v_z = d t_1$
and using Eq.\,\eqref{EQ:3_8}.

\par
It is straightforward to derive the energy loss when the charged particle is
elastically reflected from the Si surface (at $z_1 = 0$) of the SiC substrate,
yielding

\begin{equation}
\mathcal{P}
\left({\omega,\mathbf{v}}\right) =
\frac{4(Z e)^2 }{\left({2 \pi}\right)^3\epsilon_0}
\Im \texttt{m}
\int \limits_{-\infty}^{\infty} d z_1 \int \limits_{-\infty}^{\infty} d z_2
\int d^2 \qq{}\
\epsilon^{-1} \left({z_1,z_2; \qq{},\omega}\right)
\frac{\cos\left({ \left({\omega - \mathbf{q}_{\parallel}
\cdot \mathbf{v}_{\parallel}}\right)(z_2-z_1)/v_z}\right)}
{\left({q_{\parallel} v_z}\right)^2 +
\left({\omega - \mathbf{q}_\parallel}
\cdot \mathbf{v}_{\parallel}\right)^2} \ .
\end{equation}

\par
For multiple graphene layers separated by distance $d$, we can assume
wave function localization on the layers, and neglect their overlap. Therefore, we
obtain\,\cite{Kushwaha-2001,EELS_SSC}

\begin{equation}
\label{EQ:3_15_2}
\epsilon^{-1}\left({z_1,z_2;q_{\parallel},\omega)}\right)=
\delta\left({z_1 -z_2}\right) +
\sum \limits_{j,j^\prime}
v_q \texttt{e}^{-q_{\parallel}\vert{z_1 - j d }\vert}
\Pi^{(0)}_{j} \left({
\delta_{j,j^\prime} - \Pi^{(0)}_{j} v_q \texttt{e}^{-q_{\parallel} \vert{j-j^\prime }\vert d}
}\right)^{-1}
\delta\left({z_2 - j^\prime  d}\right)\ ,
\end{equation}
where the 2D Fourier transform of the Coulomb  interaction has been introduced
by $v_q =2 \pi e^2/\epsilon_s q_{\parallel}$ with
$\epsilon_s=4\pi\epsilon_0\epsilon_b$ for background dielectric
constant $\epsilon_b$. The arguments $\left({q_{\parallel},\omega}\right)$
of the noninteracting polarization function $\Pi^{(0)}_j$ on the
$j^{\text{th}}$ layer have been omitted.
Equation\,\eqref{EQ:3_15_2} can be substituted into Eq.\,\eqref{EQ:3_15_1}
and, after calculating the integrals over $z_1$ and $z_2$, we obtain:

\begin{gather}
\label{EQ:3_17}
\left[\begin{array}{c}
\mathcal{P}_{\mathcal{T}}\left({\omega,\,\mathbf{v}}\right)\\
\mathcal{P}_{\mathcal{R}}\left({\omega,\,\mathbf{v}}\right)
\end{array}\right]=
\frac{4(Z e)^2 }{\left({2 \pi}\right)^3\epsilon_0}
\Im \texttt{m}
\sum \limits_{j,j^\prime }
\int d^2 \qq{}\
\frac{v_z^2\,v_q\,\Pi_{j,j^\prime }}
{\left[{\left({q_{\parallel} v_z}\right)^2 +
\left({\omega - \mathbf{q}_\parallel}
\cdot \mathbf{v}_{\parallel}\right)^2}\right]^2}
\left[\begin{array}{c}
\mathcal{T}\\
\mathcal{R}
\end{array}\right]\ .
\end{gather}
Additionally, we have introduced auxiliary expressions for reflective
$\mathcal{R}$ and transmissive $\mathcal{T}$ particle trajectories:

\begin{eqnarray}
\mathcal{T}& = &
q_{\parallel} \exp\left[{- i q_z (j-j^\prime ) d}\right]\ ,
\\
\mathcal{R}& = &
2 q_{\parallel} \cos \left[{q_z (j-j^\prime ) d}\right] +
\texttt{e}^{- q_{\parallel} jd}
\left[{q_z \texttt{sin}(q_z j^\prime d)- q_{\parallel} \cos(q_z j^\prime d)}\right]\ ,
\end{eqnarray}
as well as
the random-phase approximation (RPA) polarization matrix elements:

\begin{equation}
\label{EQ:3_18}
\Pi_{j,j^\prime} = \Pi_{j^\prime }^{(0)}
\left({
\delta_{j,j^\prime}
- v_q\,\Pi^{(0)}_{j}\,\texttt{e}^{-q_{\parallel} \vert{j - j^\prime}\vert d}
}\right)^{-1}\ .
\end{equation}

\par
For either a single (upper) or double (lower) layer configuration, Eq.\,\eqref{EQ:3_17}
assumes the form

\[
\mathcal{P}_{\mathcal{T}(\mathcal{R})}\left({\omega,\,\mathbf{v}}\right)
=\frac{4(Z e)^2 }{\left({2 \pi}\right)^3\epsilon_0}
\Im \texttt{m}
\int
\frac{v_z^2\,d^2 \qq{}}
{\left[{\left({q_{\parallel} v_z}\right)^2 +
\left({\omega - \mathbf{q}_\parallel}
\cdot \mathbf{v}_{\parallel}\right)^2}\right]^2}
\]
\begin{equation}
\label{EQ:3_19}
\times\left\{
\begin{array}{c}
q_{\parallel}\,v_q \Pi^{(0)}_{1}\left[1-v_q \Pi^{(0)}_{1}\right]^{-1}\\
\\
\mathcal{T}(\mathcal{R})\left[
\left({1-v_q \pp{1}}\right) \left({1-v_q \pp{2}}\right)-v^2_q\,\pp{1}\pp{2}\,\texttt{e}^{-2q_{\parallel} d}\right]^{-1}
\end{array}\right.\ .
\end{equation}
Here, we have redefined the transmitted (reflected) notations as

\begin{gather}
\label{EQ:3_20}
\mathcal{T} = 2q_{\parallel}\,v^2_q\,\texttt{e}^{- q_{\parallel}d}\,
\cos(q_zd)\,
\pp{1} \pp{2}
+ v_q\left({\pp{1} + \pp{2} - 2 v_q\,\pp{1} \pp{2}}\right)\ ,\\
\label{EQ:3_21}
\notag
\\
\mathcal{R} =\texttt{e}^{-3q_{\parallel}d}\,v_q\left(
\pp{2}\,\texttt{e}^{q_{\parallel}d}\,
\left[2 q_{\parallel}\,\texttt{e}^{2q_{\parallel}d}+ q_z\,
\texttt{sin}(2q_zd) -q_{\parallel}\,\cos(2q_zd)\right]\right.
\\
\notag
+\pp{1}\left\{
v_q\,\pp{2}\left[q_z\left(1-\texttt{e}^{2q_{\parallel}d}\right)
\texttt{sin}(q_zd)+q_{\parallel}
\left({5\texttt{e}^{2q_{\parallel}d}-1}\right)
\cos(q_zd)-4q_{\parallel}\,\texttt{e}^{3q_{\parallel}d}\right]\right.\\
\notag
\left.\left.+\texttt{e}^{2q_{\parallel}d}\left[
2q_{\parallel}\,\texttt{e}^{q_{\parallel}d}+ q_z\,\texttt{sin}(q_zd)
-q_{\parallel}\,\cos(q_zd)\right]
\right\}\right)\ .
\end{gather}

\section{Energy loss in MLG}
\label{SEC:3}

Let us consider MLG with the $SiC(0001)$ substrate lying in the  $z=0$ plane.
We shall ignore contribution to EELS from $\sigma-$electrons,
substrate surface imperfections and its optical (FK-) phonons\,\cite{LIU:2010}
as well as the resonant plasmon coupling with single-particle excitations.\,\cite{TEGENKAMP:2011}
For simplicity, we shall limit ourselves to only one or two graphene layers located
at $z = d$ and $z=2 d$. In accordance with the classification in Ref.\,\onlinecite{LU:2009}
those are referred to as 0ML and 1ML correspondingly.
One of the effects of the substrate is $n-$doping of the graphene layers,
thus causing the chemical potential $\mu_{i} >0$, where $i=0,1$ labels the layers.
Hereafter we presume that both layers are held at the same chemical potential $\mu$.
The chemical  potential $\mu$  is related to the  Fermi wave vector $k_F = \mu/\hbar v_F$.
The other effect is more subtle, and affects only zero layer.
This layer, often referred to as a "buffer layer",
exhibits crossover from Dirac to conventional 2DEG by virtue of the substrate induced
gap $E_g$. Ultimately,
this gap is related to the symmetry breaking between $A$ and $B$ sublattices of the buffer
graphene layer. This effect depends on two main factors, namely,
the angle between the $\Gamma - K$ line and the principal axis of the $\texttt{SiC}$ substrate
and the degree of the hydrogen passivation of the substrate.
Formally, the $\pi-$electron dispersion around the $K\ (K^\prime)$ point  becomes

\begin{equation}
\label{EQ:2_1}
E_{k} = \sqrt{\left({\hbar v_F k}\right)^2+\left({E_{g}/2}\right)^2} \ ,
\end{equation}
where $v_F$ is the Fermi velocity for free standing graphene. The effect of the substrate
on the other graphene layers is mitigated by the buffer layer and the electrons
obey conventional Dirac dispersion ($E_g = 0$ in Eq.\,\eqref{EQ:2_1}).
Since it plays a crucial role in EELS\,\eqref{EQ:3_19}, we
give the full form of noninteracting polarization along the real frequency axis:

\begin{gather}
\label{EQ:POLARIZATION}
\Pi^{(0)}_{j}(q,\omega + i 0^+) = -\frac{2 \mu}{\pi \hbar^2 v^2_F} +
\frac{q^2}{4 \pi \sqrt{\vert{\hbar^2 v_F^2 q^2 - \hbar^2 \omega^2}\vert}}
\\
\notag
\times\left\{\left[iG_>(x_{1,-})-iG_>(x_{1,+})\right]1_<
+\left[G_<(x_{1,-})+iG_>(x_{1,+})\right]2_<\right.\\
\notag
+\left[G_<(x_{1,+})+G_<(x_{1,-})\right]3_<
+\left[G_<(x_{1,-})-G_<(x_{1,+})\right]4_<
+\left[G_>(x_{1,+})-G_>(x_{1,-})\right]1_>\\
\notag
+\left[G_>(x_{1,+})+iG_<(x_{1,-})\right]2_>
+\left[G_>(x_{1,+})-G_>(-x_{1,-})-i\pi(2-x_0^2)\right]3_>\\
\notag
+\left.\left[G_>(-x_{1,-})+G_>(x_{1,+})-i\pi(2-x_0^2)\right]4_>
+\left[G_0(x_{1,+})-G_0(x_{1,-})\right]5_>\right\}\ .
\end{gather}
Here, the following notations for the region functions have been introduced:

\begin{eqnarray*}
x_0&=&\sqrt{1+\frac{E_g^2}{\hbar^2 v_F^2 q^2-\hbar^2\omega^2}}\ ,\nonumber\\
x_{1,\pm}&=&\frac{2 \mu \pm\hbar \omega}{\hbar v_F q}\ ,\nonumber\\
x_{2,\pm}&=&\sqrt{\hbar^2 v_F^2 (q\pm k_F)^2 + (E_g/2)^2}\ ,\nonumber\\
x_{3}&=&\sqrt{\hbar^2 v^2_F q^2 + E^2_g}\ ,\nonumber\\
G_<(x)&=&x\sqrt{x_0^2-x^2}-(2-x_0^2)\cos^{-1}(x/x_0)\ ,\nonumber\\
G_>(x)&=&x\sqrt{x^2-x_0^2}-(2-x_0^2)\cosh^{-1}(x/x_0)\ ,\nonumber\\
G_0(x)&=&x\sqrt{x^2-x_0^2}-(2-x_0^2)\sinh^{-1}(x/\sqrt{-x_0^2})\ ,
\end{eqnarray*}
with the corresponding regions defined as

\begin{eqnarray*}
1_< = &\theta \left({ \mu - x_{2,-} - \hbar \omega }\right)\ ,\\
2_< = &\theta \left({ - \hbar \omega - \mu + x_{2,-} }\right)
\theta \left({ \hbar \omega + \mu - x_{2,-} }\right)  \theta \left({ \mu + x_{2,+} - \hbar \omega }\right)\ ,\\
3_< = &\theta \left({ -\mu + x_{2,-} - \hbar \omega }\right)\ ,\\
4_< = &\theta \left({ \hbar \omega + \mu - x_{2,+} }\right)  \theta \left({\hbar v_F q - \hbar \omega }\right)\ ,\\
1_> =  &\theta \left({ 2 k_F - q }\right)  \theta \left({ \hbar \omega - x_3 }\right)
\theta \left({\mu + x_{2,-} - \hbar \omega }\right)\ ,\\
2_> =  &\theta \left({ \hbar \omega - \mu - x_{2,-} }\right)
\theta \left({ \mu + x_{2,+} - \hbar \omega }\right)\ ,\\
3_> =  &\theta \left({ \hbar \omega - \mu - x_{2,+} }\right)\\
4_> =  &\theta \left({q - 2 k_F }\right)  \theta \left({ \hbar \omega - x_3 }\right)
\theta \left({ \mu +x_{2,-} - \hbar \omega }\right)\ ,\\
5_> =  &\theta \left({ \hbar \omega - \hbar v_F q}\right) \theta \left({x_3 - \hbar \omega}\right)\ .
\end{eqnarray*}

Given Eq.\,\eqref{EQ:POLARIZATION}, the plasmon dispersion relation
can be obtained as zeros of the real part of Eq.\,\eqref{EQ:3_15_2}.
Alternatively, those solutions are given by the poles of the imaginary
part of the RPA polarization in Eq.\,\eqref{EQ:3_18}. The regions of
$(\omega,q)$ space of non-zero imaginary part of non-interacting polarization
\eqref{EQ:POLARIZATION} are referred to as particle-hole continuum.
The plasmon branches in the particle-hole regions are Landau damped,
and if excited rapidly lose their energy to the single-particle excitations.

\begin{figure}[htbp]
\begin{center}
\includegraphics*[width=8cm]{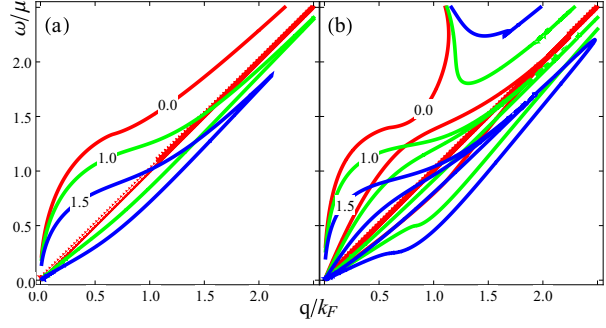}
\caption{(Color online) Undamped (two higher frequency branches) and damped (two lower in frequency branches)
plasmon dispersion relations for gapped graphene.  The curves in  (a)
correspond to  0ML(single layer),  the curves in (b) to 1ML(double layer) configuration. The red,
green and blue curves show the plasmon dispersion for $E_g/\mu = \left\{ {0.0,1.0, 1.5}\right\}$,
respectively.}
\label{FIG:2}
\end{center}
\end{figure}

The plasmon branches for $ k_F d =1$ and several values of $E_g$ are presented
in Fig.\,\ref{FIG:2}. In the long-wavelength limit the single graphene layer (0ML) exhibits
a single undamped plasmon branch [Fig.\,\ref{FIG:2}(a)] given by

\begin{equation}
\label{EQ:2_2}
\omega_p^2 = q \mathcal{D}(E_g)\ ,
\end{equation}
where the plasmon Drude factor is defined as

\begin{equation}
\label{EQ:2_3}
\mathcal{D}(E_g) = \frac{2 \mu e^2}{\hbar^2 \epsilon_{\texttt{SiC}}} \left[{1-(E_g/2\mu)^2}\right]\ .
\end{equation}
In the double layer configuration (1ML), there are two plasmon branches, i.e.,
the symmetric  $\omega_+$ and asymmetric  $\omega_-$ modes.
For large interlayer distance $ k_F d \gg 1$, the two branches are qualitatively similar
and given by

\begin{gather}
\label{EQ:2_4}
\omega^2_+ = q \mathcal{D}(0)\ ,\\
\notag
\omega^2_- = q \mathcal{D}(E_g) \ .
\end{gather}
In the opposite limit of closely placed layers with $ k_F d \ll 1$,  the asymmetric
branch becomes acoustic with frequency defined by

\begin{equation}
\label{EQ:2_5}
\omega^2_- = 2 d q^2 \mathcal{D}(E_g) \ .
\end{equation}
We note that the asymmetric branch is always smaller in frequency than symmetric mode.
In the next section, we shall combine Eqs.\,\eqref{EQ:3_19}, \eqref{EQ:3_20} with \eqref{EQ:POLARIZATION} and
simulate the energy loss for two distinct  cases, namely when the charged
particle motion is parallel  and perpendicular to the graphene layers.

\section{Numerical results and Discussion}
\label{SEC:4}

We begin our discussion with the energy loss rates in Eq.\,\eqref{EQ:3_12} for
single and symmetric double layer configurations. Our  energy loss results
for charged particle motion parallel to the surface are shown in Fig.\,\ref{FIG:1}.
The separation between the two layers was
chosen as  $ k_Fd = 1.0$.  The charged particle travels above the
graphene surface at a height $ k_F z_1= 2.0$. All  frequencies
were analytically continued to the complex plane and acquire a  small
positive loss rate $\omega \rightarrow \omega + i \gamma$ with $\gamma/\mu = 10^{-4}$.
We notice that the plasmon contribution becomes smaller with increasing energy gap  $E_g$.
This can be easily attributed to decreasing plasmon Drude factor $\mathcal{D}(E_g)$.
The contribution from the particle-hole modes is largely unaffected by the gap.
In the double layer configuration, a  small spike at the beginning
of the plasmon contribution region is due to the two branches.
Both undamped optical and acoustic  plasmon branches contribute. However, with
 increased  particle velocity, only the optical branch contributes,
as it is demonstrated in  Fig.\,\ref{FIG:2}.  Since the spectral weight of the optical
branch  is larger than that of the acoustic branch,  we see that the spike appears.
We note that one can obtain a closed-form  analytic  expression for the loss rate if
we include  only the long wavelength plasmon contributions.\,\cite{fessatidis} In any
event, this must be  the leading contribution
for small and moderate particle velocities $v_{\parallel}/v_F <6$, because of  the
$\texttt{e}^{- q z_1}/q$ factor.

\par
We now consider a single, free-standing  $(E_g =0)$ and double-layer
graphene    and present results for  the energy loss spectra of a charged particle
moving  perpendicular to the graphene surface. By symmetry,  the transmission
and reflection spectra from the single layer must be identical.
The transmission spectra are shown in Fig.\,\ref{FIG:3}.
The energy loss was calculated using either the full version of the noninteracting
polarization function in the RPA \eqref{EQ:POLARIZATION} shown in Fig.\,\ref{FIG:3}(a),
or its long-wavelength plasmon pole approximation
[Eqs.\,\eqref{EQ:2_2} and \eqref{EQ:2_4}] as seen in Fig.\,\ref{FIG:3}(b).
One may classify three different scattering regimes depending on the
charged particle velocity $v_z/v_F$. In the low-velocity regime, as shown in
column (1) of Fig.\,\ref{FIG:3} for single and double layers, the particle-hole
excitations dominate the scattering.
There is an absorption spike due to the Landau damped
plasmon mode which separates the particle-hole  continuum from the undamped
plasmon modes as shown in Fig.\,\ref{FIG:2}(a).
The strength of the spike diminishes with increased charged particle speed.
The position of such
resonance absorption at $(\approx 1.5)$ is independent of $v_z$,
indicating the effect due to linear dispersion of the damped plasmon branch.
In the intermediate velocity regime given in columns (2) and (3) of
Fig.\,\ref{FIG:3},  the damped and undamped plasmon mode contributions
are comparable.
However, in the high-velocity limit, shown in columns (4)
of Fig.\,\ref{FIG:3}, the undamped plasmons dominate the spectra.
In that regime, one can employ the polarization in its long wave limit
to a good approximation, as we can see in  Fig.\,\ref{FIG:3}(b).
The approximation works for both the single
and double-layer configuration.
In the work of Allison, et al.,\,\cite{Miskovic}
an attempt was made to   compensate for the red shift of the absorption
maximum by introducing a restoring force in the electron liquid. However,
 the origin of this restoring force was not clear.
Additionally,  the approximation introduced in their work
does not yield results which  converge on those when the RPA
polarization for graphene is employed.
Another observation is that the
double layer just doubles the absorption of the single layer,
without any perceptible spectral shift.
This is a consequence of large interlayer distance.
When the layers are closely packed the absorption peak is blue shifted
with increasing number of layers,
in agreement with Ref.\,\onlinecite{LU:2009}

\par
In Fig.\,\ref{FIG:4}, we present the transmission and reflection spectra for epitaxial
MLG. The zeroth layer only acquires the energy gap.
This gap increases in columns (1) through (3).
Our principal observation is that in the
intermediate velocity regime, given by columns (2) and (3) in
Fig.\,\ref{FIG:4},  the absorption spectrum splits into two peaks.
The one identified with  the damped plasmon peak ($\hbar\omega/\mu \approx 1.5$)
comes from the upper graphene layer for which $E_g = 0$.
The other peak
may be attributed to the symmetry-broken zeroth layer absorption ($E_g \ne 0$).
We note that in the high velocity limit, shown in column (4) of
Fig.\,\ref{FIG:4},  the gap results in a red shift of the absorption maximum.
The small splitting of the peak is also visible on the damped plasmon contribution.
As a matter of fact, the layers are so far apart that one cannot identify
a plasmon mode as  the acoustic branch.
The reflection spectrum qualitatively mimics the transmitted spectrum, but
doubles in height. This is because the charged particle spends twice as much
time in between the layers compared to the transmissive case. Of course, the
particle expends most of its energy on this part of the trajectory.

\par
To see the acoustic branch (Fig.\,\ref{FIG:2}(b)) in the spectra, we
vary intralayer distance in (rows (a) through (c) of Fig.\,\ref{FIG:5}.
for all  velocity regimes [panels (1)-(4)].
Several values of the gap are also shown on the graph.
The low-frequency acoustic branch boosts the long wavelength absorption
($\hbar\omega/\mu \approx 0$).
This is especially so for small energy gap, since the separation between
the branches is well pronounced [Fig.\,\ref{FIG:2}(b)].
In the high velocity regime or result os small interlayer separation
[Fig.\,\ref{FIG:5} (a.4)] ore result
is qualitatively similar to Fig.3 of Ref.\,\onlinecite{LU:2009}.
With increasing angle of incidence
$\theta_i = \texttt{tan}^{-1}(v_{\parallel}/v_z)$ the sharp peak of the particle hole absorption
becomes less pronounced and finally reaches
\footnote{in the sense of $\omega \sim  v_{\parallel}\,\texttt{sin}\theta_i$.}
the smooth curve as in Fig.\,\ref{FIG:1}(a.1- a.4).
The plasmon peak is blue shifted with $\texttt{sin} \theta_i$ finally reaching those of Fig.\,\ref{FIG:1}(a.5).
Since the plasmon momentum is proportional to $\texttt{sin} \theta_i$ one can interpret this blue shift as the Raman shift due to plasmon activation.
\footnote{In our theory, we force the particle on the prescribed trajectory, thus making the incidence and the scattering angle to be the same.
Therefore, Eq.\,(\ref{EQ:3_1}) of Ref.\,22 confirms that the plasmon momentum $q$ is proportional to the sine of the incidence angle.}
Given its linear nature we deuce that for small interlayer distance mostly acoustic plasmon branch is activated,
thus explaining Fig.2(d) of Ref.\,\onlinecite{LU:2009}.
This interpretation is further confirmed by the comparison with the large interlayer separation Raman shift,
which follows $\sqrt{\texttt{sin}\theta_i}$ pattern.
Unfortunately, as it follows from Fig.\,\ref{FIG:2}(b), it is hard to deduce existence of the gap from the acoustic plasmon branch.
Experimentally one would have to transfer MLG onto
a non-polar substrate and observe the increase in the plasmon slope if the gap was originally present.

\par
Comparing the plasmon
absorption in the parallel and perpendicular cases, the plasmons contribute
more to  the former in the low-velocity regime, whereas in the latter case,
their contribution is most pronounced for higher incoming charged particle velocities.

\section{Summary}
\label{SEC:5}

We have investigated the role played by a gap in the   energy dispersion
on the absorption spectra of single and double layer configurations.
All velocity regimes for the external charged particle moving perpendicular to the graphene surface
were reported in Figs.\,\ref{FIG:3},\,\ref{FIG:4} and \ref{FIG:5}.  The plasmon pole
approximation for the polarization  agrees well with the results obtained
with the full polarization in the RPA in the high-velocity regime only.
The speed of the external charged particle determines whether the
plasmon or particle-hole excitations dominate the scattering.
Consequently, since gapped graphene has a different plasma excitation spectrum than
free-standing graphene, its stopping power may carry distinct signatures of the substrate induced gap.
We also demonstrated that our formalism can qualitatively describe experimental data of ELLS.
It also allowed us to interpret the observed linear plasmon dispersion as coming from the acoustical undamped branch.

\section*{Acknowledgement(s)}

This research was supported by  contract \# FA 9453-07-C-0207 of AFRL. DH would also like to thank Prof. Xiang Zhang for hosting the Visiting
Scientist Program sponsored by AFOSR.

\appendix

\section{An alternative derivation of the energy loss formula}
\label{AP:1}

Here, we derive Eq.\eqref{EQ:3_15_1}  by a method similar to that in
Refs.\ \onlinecite{Mills} and \onlinecite{Miskovic}. First, we introduce
the 2D Fourier transform and its inverse given by

\begin{gather}
\label{EQA:3_16}
\left[{\mathcal{F}_{2D,\alpha} f }\right] (\mathbf{q}_{\alpha \parallel}, z_{\alpha},\omega_{\alpha}) =
\int \limits_{-\infty}^{\infty}  d t_{\alpha}\
\texttt{e}^{i \omega_{\alpha} t_{\alpha}}
\int d^{2} \mathbf{r}_{\alpha \parallel} \
\texttt{e}^{- i \mathbf{q}_{\alpha,\parallel} \cdot \mathbf{r}_{\alpha \parallel}}
f\left({\mathbf{r}_{\alpha \parallel}, z_{\alpha},t_{\alpha}}\right)\ ,\\
\label{EQA:3_16}
\left[{\mathcal{F}^{-1}_{2D,\alpha} f }\right] (\mathbf{r}_{\alpha \parallel}, z_{\alpha},t_{\alpha}) =
\frac{1}{\left({2 \pi}\right)^3}
\int \limits_{-\infty}^{\infty}  d \omega_{\alpha} \texttt{e}^{-i \omega_{\alpha} t_{\alpha}}
\int d^{2} \mathbf{q}_{\alpha \parallel} \texttt{e}^{ i \mathbf{q}_{\alpha \parallel} \cdot \mathbf{r}_{\alpha \parallel}}
f\left({\mathbf{q}_{\alpha \parallel}, z_{\alpha},\omega_{\alpha}}\right)\ ,
\end{gather}
where we have adopted the notation of Camley and Mills\,\cite{Mills}
with $\alpha = 1,2$. By applying such a Fourier transformation defined
in \eqref{EQA:3_16} to the Poisson equation \eqref{EQ:3_3}, we  obtain after
some straightforward algebra the following differential equation

\begin{equation}
\left({\frac{\partial^2}{\partial z^2_2} - q^2_{2\parallel}}\right)
\Phi_{\texttt{ext}} \left({ \mathbf{q}_{2\parallel}, z_2, \omega_2}\right) =
-\frac{Z e}{v_z\epsilon_0}\,\texttt{e}^{i \left({\omega_2 - \mathbf{q}_{2\parallel}
\cdot \mathbf{v}_{\parallel}}\right) z_2/v_z} \ .
\end{equation}
By applying the boundary conditions $\Phi_{\texttt{ext}}\left({\mathbf{q}_{2\parallel}, \pm \infty, \omega_2}\right) = 0$,
the solution may be written as

\begin{equation}
\Phi_{\texttt{ext}}\left({\mathbf{q}_{2\parallel}, z_2, \omega_2}\right) =
\frac{Z e\,v_z\,\texttt{e}^{i \left({\omega_2 - \mathbf{q}_{2\parallel}
\cdot \mathbf{v}_{\parallel}}\right) z_2/v_z} }
{\epsilon_0\left[\left({q_{2\parallel} v_z}\right)^2 +
\left({\omega_2 - \mathbf{q}_{2\parallel} \cdot \mathbf{v}_{\parallel}}\right)^2\right]}\ .
\end{equation}
Therefore, the external potential assumes the form

\begin{gather}
\label{EQA:3_17}
\Phi_{\texttt{ext}}\left({\mathbf{r}_{2\parallel}, z_2, t_2}\right) =
\mathcal{F}^{-1}_{2d,2} \Phi_{\texttt{ext}}\left({\mathbf{q}_{2\parallel}, z_2, \omega_2}\right)  \\
\notag
=
\frac{Z e\,v_z}{\left({2 \pi}\right)^3\epsilon_0}
\int \limits_{-\infty}^{\infty}  d \omega_{2}\,\texttt{e}^{-i \omega_{2} t_{2}}
\int d^{2} \mathbf{q}_{2 \parallel}\
\frac{\exp\left({ i \mathbf{q}_{2\parallel} \cdot \mathbf{r}_{2 \parallel}
+ i \left({\omega_2 - \mathbf{q}_{2\parallel}
\cdot \mathbf{v}_{\parallel}}\right) z_2/v_z}\right)}
{\left({q_{2\parallel} v_z}\right)^2 +
\left({\omega_2 - \mathbf{q}_{2\parallel} \cdot \mathbf{v}_{\parallel}}\right)^2}\ .
\end{gather}
Similarly,   we  make a Fourier representation for the nonlocal
inverse dielectric function as

\begin{gather}
\label{EQA:3_18}
\epsilon^{-1} \left({\mathbf{r}_{1\parallel},z_1,t_1; \mathbf{r}_{2\parallel},z_2,t_2}\right)=
\epsilon^{-1} \left({z_1,z_2; \rr{1}-\rr{2}, t_1-t_2}\right) \\
\notag
=
\frac{1}{\left({2 \pi}\right)^3}
\int \limits_{-\infty}^{\infty} d \omega\  \texttt{e}^{-i \omega (t_1-t_2)}
\int d^2 \mathbf{q}_{\parallel}\
\texttt{e}^{i \mathbf{q}_{\parallel} \cdot (\rr{1}-\rr{2})}\,
\epsilon^{-1} \left({z_1,z_2;\qq{},\omega}\right) \ .
\end{gather}
Combining Eqs.\,\eqref{EQA:3_17} and \eqref{EQA:3_18} with Eq.\,\eqref{EQ:3_2},
we obtain

\begin{gather}
\label{EQA:3_19}
\Phi_{\texttt{tot}} \left({\rr{1}, z_1, t_1}\right) =
\frac{Z e\,v_z}{\left({2 \pi}\right)^3\epsilon_0}
\int \limits_{-\infty}^{\infty} d \omega\  \texttt{e}^{-i \omega t_1}
\int d^2 \qq{} \  \texttt{e}^{i \qq{} \cdot \rr{1}}
\int \limits_{-\infty}^{\infty} dz_2
\\
\notag
\times \epsilon^{-1} \left({z_1,z_2; \qq{},\omega}\right)
\frac{\exp\left({ i \left({\omega - \mathbf{q}_{\parallel}
\cdot \mathbf{v}_{\parallel}}\right) z_2/v_z}\right)}
{\left({q_{\parallel} v_z}\right)^2 +
\left({\omega - \mathbf{q}_\parallel}
\cdot \mathbf{v}_{\parallel}\right)^2} \ .
\end{gather}
Here, we used the following identities:

\begin{gather*}
\int \limits_{-\infty}^{\infty} dt_2\
 \texttt{e}^{-i (\omega_2 - \omega) t_2} = 2 \pi \delta \left({\omega_2 -\omega}\right)\ ,\\
\int d^2 \rr{2}\,\texttt{e}^{-i(\qq{}-\qq{2}) \cdot \rr{2}} \left({2 \pi}\right)^2
\delta \left({\qq{} -\qq{2}}\right) \ .
\end{gather*}
The force acting on the particle is given by combining Eq.\,\eqref{EQA:3_19}
and Eq.\,\eqref{EQ:3_1}, with the replacement
$\nabla_{1} = \partial/\partial \mathbf{v} t_1$.
The resulting expression is inserted into Eq.\,\eqref{EQ:3_6}, yielding

\begin{gather}
\label{EQA:3_20}
W_{\theta} \left({\mathbf{v}}\right) = \int \limits_{-\infty}^{\infty}
d (\mathbf{v}\,z_1/v_z)\cdot
\mathbf{F} \left({1}\right) =
-\frac{(Z e)^2}{\left({2 \pi}\right)^3\epsilon_0}
\int \limits_{-\infty}^{\infty} d z_1
\int \limits_{-\infty}^{\infty} dz_2
\int d^2 \qq{}
\int \limits_{-\infty}^{\infty} d \omega\  i \omega\,\texttt{e}^{-i \omega z_1/v_z}\,
\texttt{e}^{i \qq{} \cdot \mathbf{v}_{\parallel} z_1/v_z}
\\
\notag
\times
\epsilon^{-1} \left({z_1,z_2; \qq{},\omega}\right)
\frac{\exp\left({ i \left({\omega - \mathbf{q}_{\parallel}
\cdot \mathbf{v}_{\parallel}}\right) z_2/v_z}\right)}
{\left({q_{\parallel} v_z}\right)^2 +
\left({\omega - \mathbf{q}_\parallel}
\cdot \mathbf{v}_{\parallel}\right)^2} \\
\notag
=
-\frac{(Z e)^2 }{\left({2 \pi}\right)^3\epsilon_0}
\int \limits_{-\infty}^{\infty} d z_1 \int \limits_{-\infty}^{\infty} d z_2
\int d^2 \qq{}
\int \limits_{-\infty}^{\infty} d \omega\  i  \omega\,
\epsilon^{-1} \left({z_1,z_2; \qq{},\omega}\right)
\frac{\exp\left[{ i \left({\omega - \mathbf{q}_{\parallel}
\cdot \mathbf{v}_{\parallel}}\right)(z_2-z_1)/v_z}\right]}
{\left({q_{\parallel} v_z}\right)^2 +
\left({\omega - \mathbf{q}_\parallel}
\cdot \mathbf{v}_{\parallel}\right)^2} \ .
\end{gather}
Now, let us change the dummy variables $\omega \to -\omega \; \mathbf{q}_{\parallel} \rightarrow - \mathbf{q}_{\parallel}$
in the above equation and utilize the symmetry relation in Eq.\,\eqref{EQ:3_10}.
We obtain

\begin{equation}
W_{\theta} \left({\mathbf{v}}\right) =
\frac{2(Z e)^2 }{\left({2 \pi}\right)^3\epsilon_0}\
\Im \texttt{m}
\int \limits_{-\infty}^{\infty} d z_1 \int \limits_{-\infty}^{\infty} d z_2
\int d^2 \qq{}
\int \limits_{0}^{\infty} d \omega\   \omega\,
\epsilon^{-1} \left({z_1,z_2; \qq{},\omega}\right)
\frac{\exp\left[{ i \left({\omega - \mathbf{q}_{\parallel}
\cdot \mathbf{v}_{\parallel}}\right)(z_2-z_1)/v_z}\right]}
{\left({q_{\parallel} v_z}\right)^2 +
\left({\omega - \mathbf{q}_\parallel}
\cdot \mathbf{v}_{\parallel}\right)^2} \ .
\end{equation}

\begin{figure}[p]
\begin{center}
\includegraphics*[width=8cm]{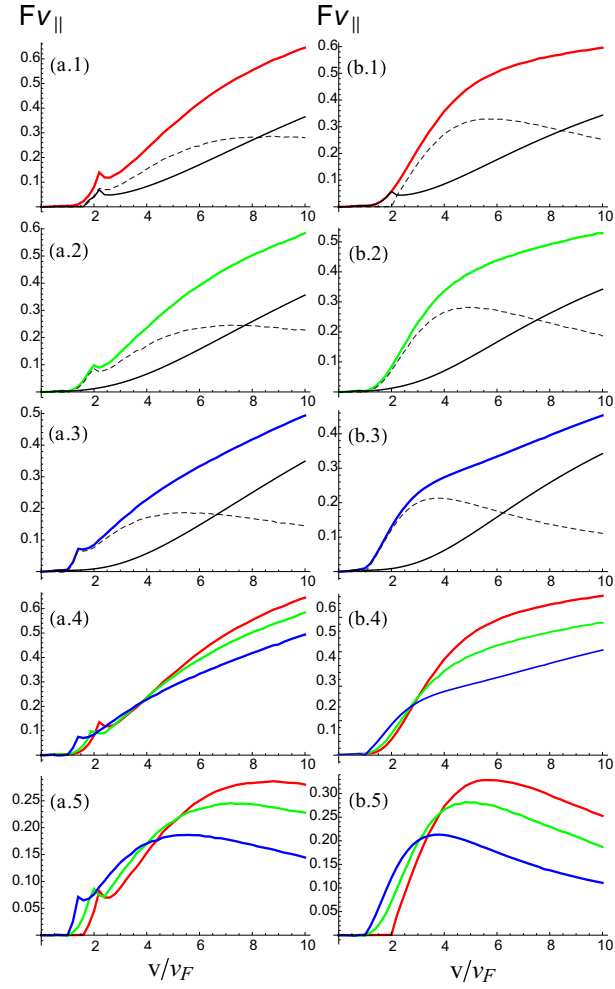}
\caption{(Color online)  Stopping power, as a function of
charged particle velocity in units of $(Z e / 2 \pi)^2 \epsilon_s^{-1}$.
Rows (a), (b) correspond to double  and single  layer configurations,
respectively. Panels (1), (2) and (3)  correspond to $E_g/\mu = \left\{ {0.0,1.0, 1.5}\right\}$
illustrated by red, green and blue curves, respectively.
The solid black curve is the particle-hole contribution, and the dashed
curve shows the plasmon  contribution. Row (4) gives the
particle-hole contribution for the three values of the gap, and row
(5) shows the associated plasmon contributions.}
\label{FIG:1}
\end{center}
\end{figure}

\begin{figure}[p]
\begin{center}
\includegraphics*[width=\textwidth]{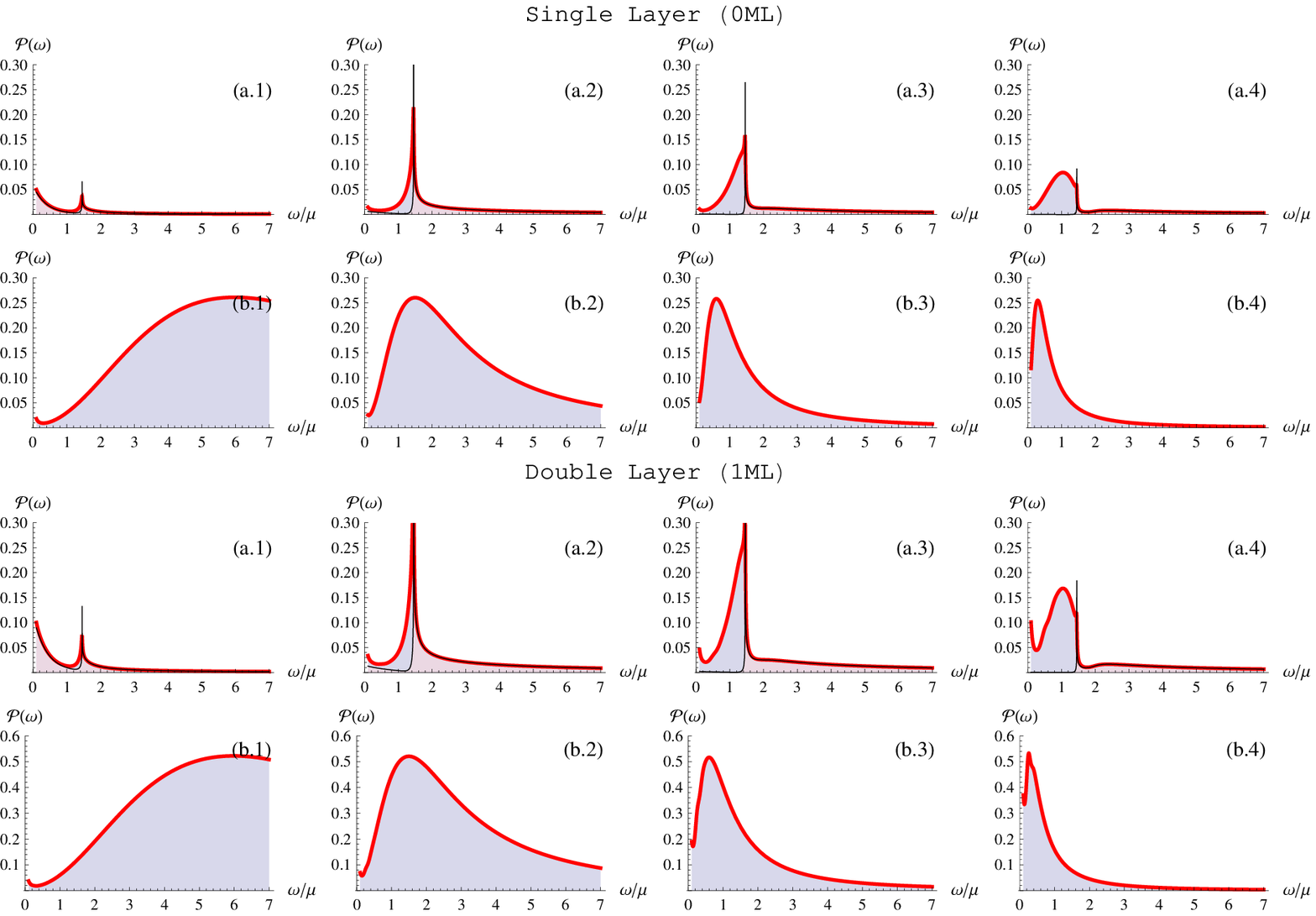}
\caption{(Color on-line)
Energy loss spectra of the charged particle transmitted through single
and double $( k_F d= 100)$ free-standing graphene. Rows (a) and (b) correspond
to the full version of the RPA polarization   and its long-wave plasmon approximation,
respectively. In columns (1) through (4), the charged particle speed increases
as  $v_z/v_F = \left\{ {0.5, 2.0, 5.0, 10.0}\right\}$.
The thin black curves correspond to  particle-hole (and damped plasmon) contributions only.}
\label{FIG:3}
\end{center}
\end{figure}

\begin{figure}[p]
\begin{center}
\includegraphics*[width=\textwidth]{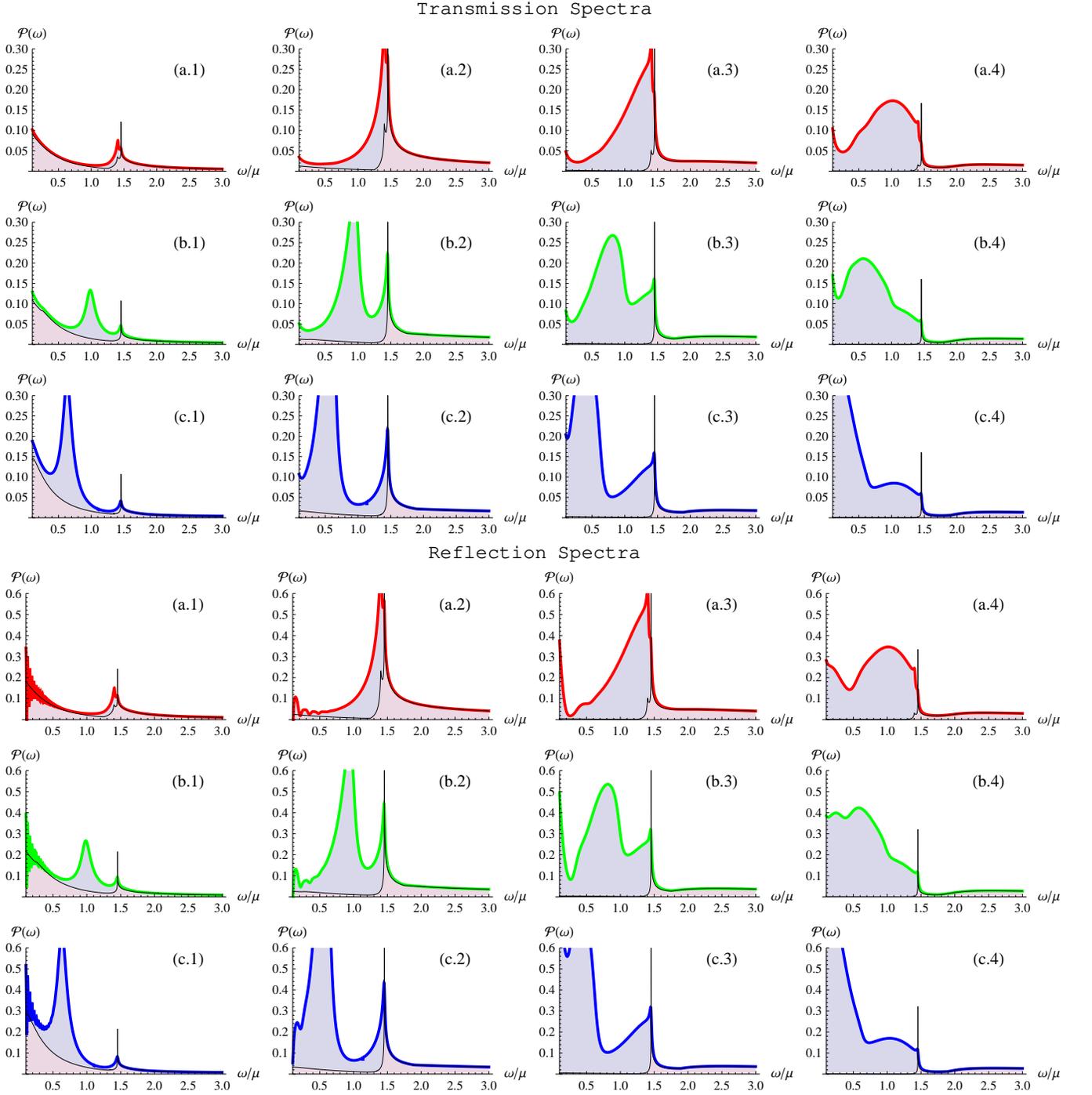}
\caption{(Color on-line)
Energy loss spectra of the charged particle transmitted through and reflected
from a double $(k_F d = 100)$ epitaxial graphene. Rows  (a) through (c)
are for increasing energy gap on the zeroth layer with
$E_g/\mu = \left\{ {0.5, 1.0, 1.5}\right\}$. In columns (1) through (4),
the charged particle speed increases as
$v_z/v_F = \left\{ {0.5, 2.0, 5.0, 10.0}\right\}$.
The thin black
curves correspond to  particle-hole (and damped plasmon) contributions only.
The color schematic correspond to Fig.\,\ref{FIG:2} and Fig.\,\ref{FIG:1}.}
\label{FIG:4}
\end{center}
\end{figure}

\begin{figure}[p]
\begin{center}
\includegraphics*[width=\textwidth]{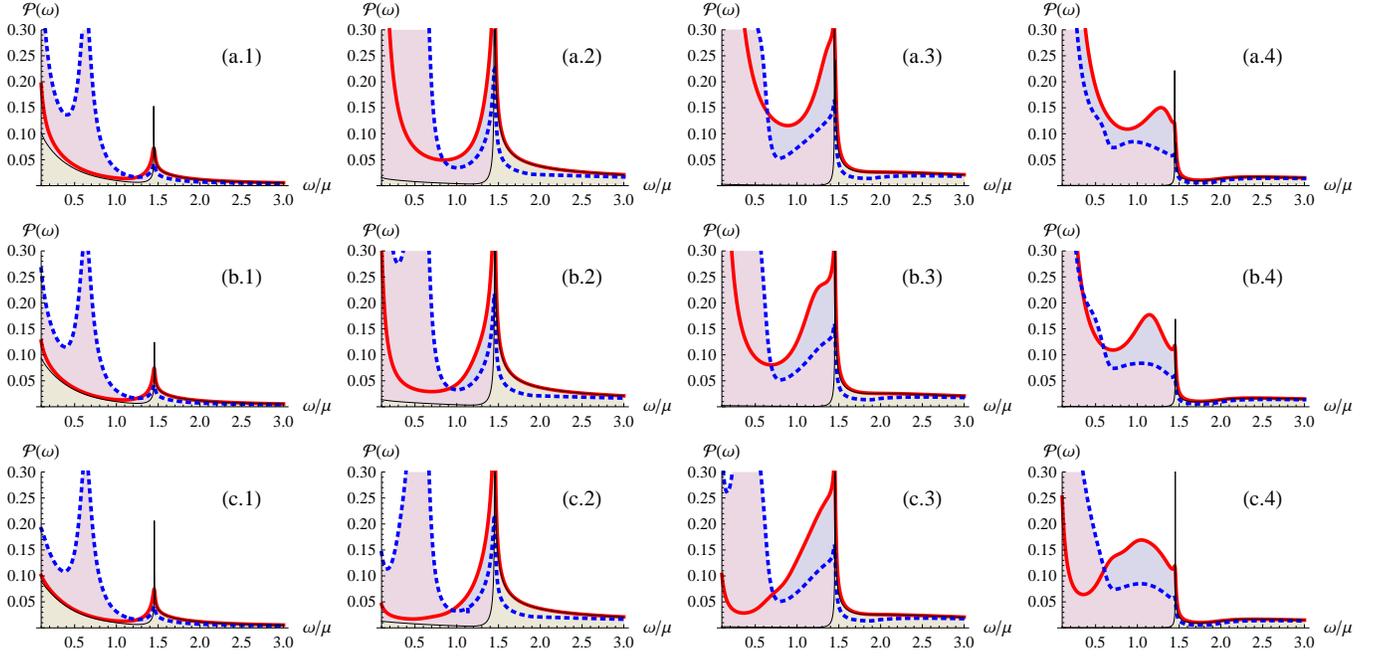}
\caption{(Color online)
Energy loss spectra of the charged particle transmitted through a double
free-standing graphene. Rows (a) through (c) correspond to increasing
interlayer distance $k_F d = \left\{ {5,10, 50}\right\}$ . In columns
(1) through  (4), the charged particle speed increases as  with
$v_z/v_F = \left\{ {0.5, 2.0, 5.0, 10.0}\right\}$. The thin black
curves correspond to  particle-hole (and damped plasmon) contributions only.
Red thick and Blue dotted curves stand for $E_g/\mu = 0.0$ and $E_g/\mu = 1.5$ correspondingly.}
\label{FIG:5}
\end{center}
\end{figure}

\end{document}